# Predicting Nanofabrication Variance: A Conditional Diffusion Model

Wenhan Zhang, Ben Mills, Oleksandr Buchnev, and Kevin F. MacDonald*
Optoelectronics Research Centre, University of Southampton, Highfield, Southampton, SO17 1BJ, UK

**ABSTRACT**. Micro- and nanostructures fabricated by focused ion beam (FIB) milling invariably deviate from input designs due to systematic and stochastic effects such as ablation-rate fluctuations, and redeposition. As such, nominally identical iterations of a given process yield a distribution of non-identical fabrication outcomes, leading to variations in the physical (optical, mechanical) properties of fabricated structures. Existing models of the milling process typically predict a single deterministic fabrication result, neglecting this variability. Here we present a conditional diffusion model that learns the statistical distribution of process outcomes conditioned on input design, enabling it to predict a realistic distribution of possible output sample structures. These predictions accurately reproduce edge morphology, linewidth variations, and defect probabilities observed in fabricated samples, enabling statistical analysis of milling performance and the identification of dominant structural and process parameter-dependent deformation modes. The work establishes a probabilistic modelling framework for fabrication variability analysis and fabrication-aware design optimization.

## I. INTRODUCTION

Focused Ion Beam (FIB) milling is a key technique in nanofabrication rapid prototyping and materials/device modification and characterisation [1-3]. It operates by raster scanning a tightly focused (usually gallium) ion beam across a substrate surface to selectively ablate material, enabling direct, locally-controlled pattern formation (including etch depth variation) in almost any metal, semiconductor, or dielectric target material. Fluctuations in beam current [4], ion–solid interactions [5], and redeposition [6] during milling introduce stochastic deviations from the desired feature geometry, and systematic artefacts related to the raster pattern or direction (relative to the orientation of design features). These deviations become increasingly significant as feature dimensions decrease, leading to local over- or under-milling, dimensional bias, and line-edge roughness, with particularly strong impact on narrow features and sharp corners. This can manifest strongly in the optically and mechanically resonant properties (frequencies/wavelengths and quality factors) of nanophotonic, plasmonic (e.g. metamaterial), and micro/nanomechanical structures (cantilevers, beams, nanostructured membranes), which are highly sensitive both to dimensional/geometric deviations from the input design and the dispersion of variations among repeated, nominally identical structural elements.

Although physics-based simulation tools such as Monte Carlo sputtering models and finite element methods [7,8] exist, they are computationally intensive, struggle to accurately model complex structures [9], and do not provide probabilistic descriptions of stochastic defects that arise during the milling process. Neural network–based approaches have emerged recently as an alternative data-based predictive framework, with generative networks enabling image-to-image mapping [10-13] from input mask designs to the output appearance of fabricated samples (in scanning electron microscope – SEM – images), using paired training data. Conditional generative adversarial networks (cGANs) [14,15], for example, require generated outputs to resemble realistic samples from the target domain, enabling mask-conditioned SEM image prediction. However, the adversarial objective forces the generator to converge towards the single most-likely output, rather than a distribution of plausible outcomes [16]. Therefore, fabrication variability cannot be characterised.

Diffusion models, on the other hand, explicitly learn the underlying data distribution and avoid mode collapse, enabling the generation of multiple physically possible outcomes rather than a single representative prediction [17]. In this work, we present a conditional diffusion model for probabilistic prediction of FIB milling outcomes conditioned on input mask geometry (Fig. 1). The model is trained on experimental paired input mask plus output SEM image datasets, to generate realistic distributions of SEM image-like realisations from a single binary mask. It reproduces key physical characteristics of milled structures and the statistical distribution of fabrication-induced variations. The predictions enable quantitative characterization of effective milling resolution and defect probability distributions, revealing orientation-dependent defect behaviours and dominant structural deformation modes.

*Contact author: kfm@orc.soton.ac.uk

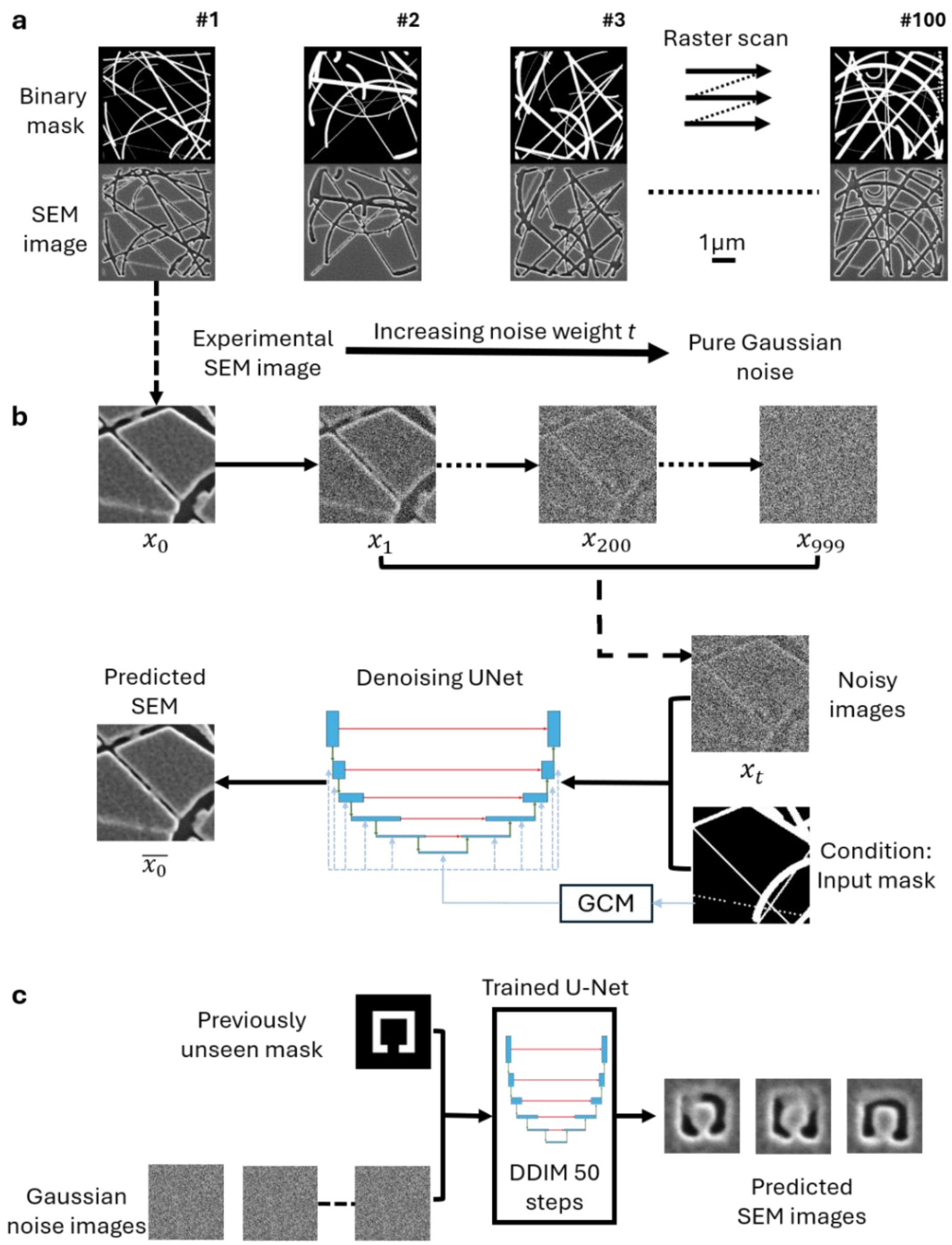

FIG 1. Conditional diffusion model for FIB milling – training and application. (a) Representative pairs of binary mask and corresponding fabricated sample SEM images used for model training and validation. Note that in the binary masks, white areas are those to be milled, while in SEM images, milled regions [exposed silicon nitride substrate] appear darker than the surrounding [gold] surface. (b) Diffusion model training procedure – the full training dataset is generated through a 'forward' process of incrementally adding Gaussian noise to sample SEM images [upper row]. A U-Net is then trained to perform the reverse 'de-noising' process, conditioned on the binary mask. (c) From a set of different Gaussian noise initializations and a previously-unseen binary mask, the trained U-Net generates an ensemble of pseudo-SEM images of [hypothetically] fabricated samples, capturing the stochastic variability of the FIB milling process.

Predictive knowledge of these variations, in advance of fabrication, opens the path to fabrication-aware design, including setting of viable fabrication tolerances, realistic estimation of feature-to-feature or device-to-device variability, and expected impact on the optical/mechanical properties of fabricated structures.

## II. CONDITIONAL DIFFUSION MODEL ARCHITECTURE AND TRAINING

We employed a conditional denoising diffusion probabilistic model [18] to map binary mask images(nanofabrication process inputs) to predicted, physically plausible SEM images of milling outcomes on a thin gold film – an archetypal target medium for nanophotonics, plasmonics, and metamaterials applications [19-21].

Figure 1a shows a representative selection of the 100 pairs of binary masks and corresponding fabricated sample SEM images used for training the diffusion model. The 830×830-pixel masks comprise random arrangements of straight and circular-segment lines of random length and linewidth, providing a diverse range of edge profiles, intersection angles, and feature orientations (with respect to the ion beam raster scan direction). Samples were fabricated and imaged using an FEI Helios Nanolab 600 DualBeam FIB/SEM system: The patterns were milled through a 50 nm thermally evaporated gold film supported on a 50 nm silicon nitride membrane, using a fixed ion beam current of 10 pA and a dose of 7 mC/cm². (Prior to FIB milling, the samples were solvent-cleaned and baked to minimise surface contamination and residual moisture.) SEM images were acquired under fixed imaging conditions of 5 kV acceleration voltage, 86 pA beam current, 25,000× magnification, 5.12 µm horizontal field width, 4.1 mm working distance, and 0° stage tilt.

In training (Fig. 1b), the diffusion model learns to reconstruct fabricated sample SEM images from versions corrupted with Gaussian noise, while being conditioned on the corresponding binary (input design) mask. The forward diffusion process comprises 999 incremental noise-addition steps, converting a randomly-selected 256×256-pixel section of each sample SEM image ($x_0$) into an almost pure-noise image ($x_{999}$). The added pattern of 8-bit noise is randomly generated at each step, differently for each of the 100 sample SEM images. A U-Net [22] is then trained, using all 100×999 corrupted images in random sequence, to perform the reverse denoising process – i.e. to recover the original SEM images. In each training iteration, the network receives an SEM image ($x_t$), the value of $t$ (i.e. information describing the amount of added noise), and the corresponding binary mask. Features are processed through four encoder and four decoder levels connected by skip connections, with a self-attention bottleneck at the coarsest resolution integrating global spatial context. The U-Net learns to predict the added noise, enabling it to reconstruct realistic SEM images based upon the input binary mask geometry.

To preserve local structural detail that would otherwise be diluted by repeated downsampling in the U-Net, a geometry-conditioning pathway is included: Spatially adaptive denormalization (SPADE) blocks [23] – small convolutional neural networks – transform the binary mask into a set of maps describing aspects of local geometry, including the proximity of neighbouring features in the horizontal and vertical directions, edge locations, and narrow-line regions. These are then applied as conditioning signals to the intermediate feature activations at every resolution level, allowing the U-Net to adapt its response according to local geometry throughout the denoising process.

The conditional diffusion model was trained using a composite loss function: The primary loss was the standard denoising mean squared error, which trains the network to predict the Gaussian noise added to each SEM image. Additional auxiliary losses were applied to improve the statistical realism of generated pseudo-SEM image outputs, particularly within fine-feature regions where fabrication variability is most pronounced. These terms compared the predicted ($\overline{x_0}$) and real ($x_0$) SEM images using spectral, histogram-based, and local-variance statistics, capturing differences in frequency content, intensity distributions, texture variation, gap-region brightness, and interior uniformity. Further geometry-aware histogram losses were included to encourage the model to reproduce the statistical distribution of defect-related intensity variations observed in geometry-dependent gap regions.

In application, given a previously-unseen binary mask and a set of independent, randomly generated Gaussian noise initialisations, the trained U-Net will then generate an ensemble of pseudo-SEM images, capturing the stochastic variability of the fabrication process (Fig. 1c). The inference process employs denoising diffusion implicit model (DDIM) sampling [24]: Unlike the original denoising diffusion probabilistic model (DDPM) sampling procedure [18], which would traverse all 999 diffusion steps defined during training, DDIM constructs images using substantially fewer denoising steps while maintaining comparable output quality – 50 in the present case, as a practical compromise between generation fidelity and computational efficiency.

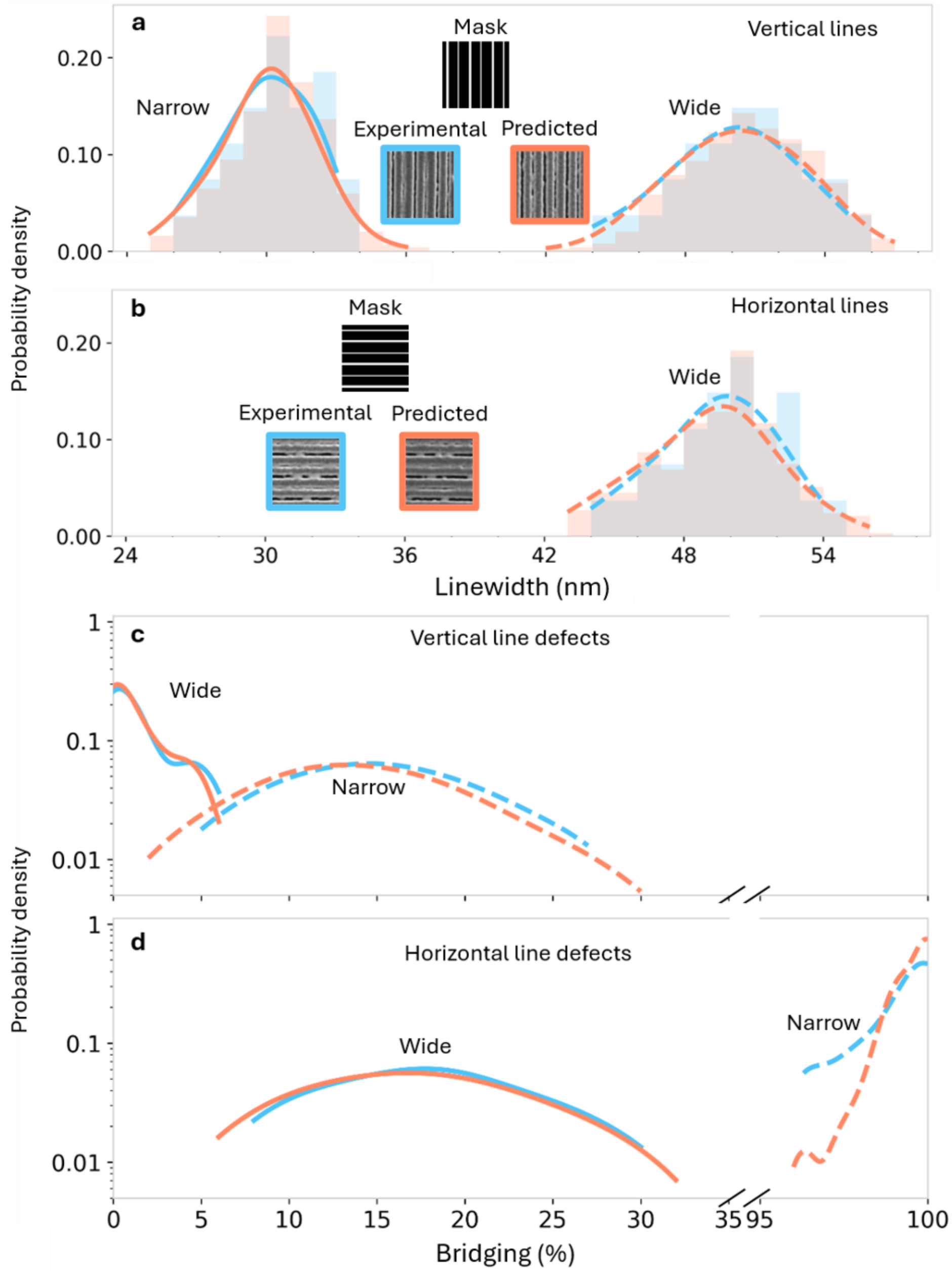

FIG 2. Diffusion model predictions vs. experimental FIB milling outcomes. (a, b) Histograms [overlaid with best-fit probability-density curves] of milled linewidths in simulated [orange bars and curves] and experimentally fabricated [blue] patterns of wide/narrow straight-line trenches milled through a gold film, in (a) vertical and (b) horizontal orientations, perpendicular and parallel respectively to the ion beam raster scan lines. Inset images show the relevant binary masks, representative section of corresponding experimental sample SEM images, and representative diffusion model-generated pseudo-SEM images. [Narrow trenches are omitted from (b) because high defect rates preclude meaningful linewidth measurement – see (d).] (c, d) Defect rates in simulated and experimentally fabricated patterns of straight-line trenches, again in (c) vertical and (d) horizontal orientations. [Linewidths are sample from binarized SEM images at ~16 nm intervals and averaged over 1 µm lengths; Defect rates are evaluated as the percentage of measurement points at which trenches are not formed.]

## III. PREDICTING FABRICATION VARIABILITY

To quantify the diffusion model's ability to predict variations in milling outcomes, we first consider the fabrication of individual straight-line trenches through the gold film – a simple geometry that enables straightforward quantitative comparison between simulated and experimental milling performance, while revealing subtle systematic effects associated with the mutual orientation of design features and the ion beam raster. We then consider the milling of square split rings – a more complex, generic plasmonic metamaterial unit-cell geometry incorporating internal and external corners, and a narrow gap of critical width – which provides for analysis of structural deformation modes in repetitive patterns. In both cases, to provide suitably challenging tests for the diffusion model, we specify relatively small feature dimensions close to the limit of FIB milling resolution, whereby relatively poor (i.e. highly variable) milling performance is expected. In all cases, experimental samples are fabricated and imaged using the same ion and electron beam parameter settings as for the training samples above.

In the first case, an 830×830-pixel binary mask pattern of parallel 5- and 9-pixel (nominally ~20 and ~36 nm) wide lines, with spacing (36 pixels, ~144 nm) >> width, was milled through the gold film, in 'vertical' and 'horizontal' orientations parallel and perpendicular respectively to the ion beam's raster scan lines. SEM images of these samples were scaled to matched 830×830-pixel size, cropped to 768×768 pixels, and divided into a set of nine contiguous, nominally identical 256×256-pixel patches, each containing three ~1 µm lengths of each linewidth (as shown inset to Figs. 2a, b). Pixel brightness in each patch each is then normalized to a 8-bit scale. From a corresponding 256×256-pixel section of the binary mask, the diffusion model was used to generate a set of 100 pseudo-SEM images in each orientation – representative examples of which are again shown inset to Figs. 2a,b.

Figure 2 compares simulated and experimental milling performance in terms of trench linewidth and the likelihood of 'bridging' defects (where the full thickness of the gold film is not removed over the full width of the intended the trench – i.e. a continuous gold bridge remains between what should be the two walls of the trench). Linewidths are evaluated for each 256-pixel length of each line (in each of the 18 experimental SEM image patches, and each of the 200 pseudo-SEM images), after binarization with a fixed intensity threshold (of 0.7), as averages over 64 equally spaced points (i.e. at 4-pixel intervals) – yielding sets of 27 experimental and 300 simulated wide/narrow linewidth measurements, in each orientation (Figs. 2a,b).

The diffusion model accurately and precisely predicts the experimentally achieved wide-trench line linewidths of 50.0±2.7 nm (vertical orientation) and 49.4±2.4 nm (horizonal), and narrow-trench linewidths of 30.0±1.8 nm (vertical only; narrow horizontal lines are not fully-formed – see below). These widths are all larger than the design widths (above) due to the finite diameter and focal depth of the ion beam, and the associated, characteristically tapered sidewall profile and rounded edges of FIB-milled features (the extent of which is a function of milling depth, i.e. film thickness).

While linewidths do not substantively depend on feature orientation, defect rates do (Fig. 2c,d) – to the extent that there is a near-complete failure to mill narrow trenches in the horizontal orientation, parallel to the ion beam raster scan lines (hence their exclusion from the analysis of linewidth above). This is because bridging defects, where gold remains over what should be exposed silicon nitride substrate, are primarily a consequence of redeposition, which depends on the sequencing of neighbouring points in the milling process. The diffusion model again accurately predicts this behaviour, yielding defect rates and distributions closely matched to those seen in experimental samples across the full range from very low (~1.3% for wide vertically oriented trenches) to very high (>99% for narrow vertically oriented trenches). The model's success in this regard is notable for the fact that the raster direction-dependent anisotropy in milling performance is not readily apparent (even with hindsight) by visual inspection of the training samples (Fig. 1a). Conventional (human-curated) FIB process development proceeds through iterative trial-and-error optimisation of the intended sample design, rather than assessment of random test patterns. However, this is time-consuming and may not immediately identify design orientation as a significant factor against the several other complex and concurrent effects of varying process parameters. By contrast, the diffusion model infers this behaviour directly from training patterns and successfully generalises it to previously unseen test geometries.

To further illustrate the diffusion model's predictive capability, in application to more complex (photonic metamaterial) geometries, we analysed the statistical variability of fabricated and simulated split-rings (Fig. 3). By design, the 180 nm square rings have a linewidth of 36 nm, and a gap width of 48 nm. A set of 128 individual split-rings were milled through the gold film and imaged as above, and from the same binary mask the diffusion model was employed to generate a set of 300 pseudo-SEM images.

Representative examples of each are shown alongside the mask in Fig. 3a.

A principal component analysis (PCA) was performed separately on the fabricated sample and diffusion-model-generated image sets to identify the dominant structural deformation modes underlying variations in fabricated/predicted sample geometry. The diffusion model yields the same sequence of deformation modes, with closely matched explained variance (EV – defined as the proportion of total variance accounted for by a given component).

This is illustrated in Figs. 3b,c for the first five principal components, which together account for around two-thirds of total variance: The first mode corresponds to linewidth variation and alone accounts for about a third of total variance (~3× as much as the next most significant component). The second and third modes capture horizontal and vertical asymmetries in the split-ring geometry, with EV ratios of ~1/8 and ~1/12, respectively – the difference in magnitude between these two being indicative of the raster-dependent anisotropy in milling performance discussed above (Fig. 2). Higher-order modes (each with EV ratios <1/20) describe progressively more localised geometric variations, including edge rounding and corner sharpness. Figure 3c shows how the spatial distributions of deformation are accurately reproduced by the diffusion model, at least for the first three principal components (which cumulatively account for around 3/5 of total geometric variance); the correlation weakens for higher-order modes representing increasingly subtle and less-frequently occurring perturbations.

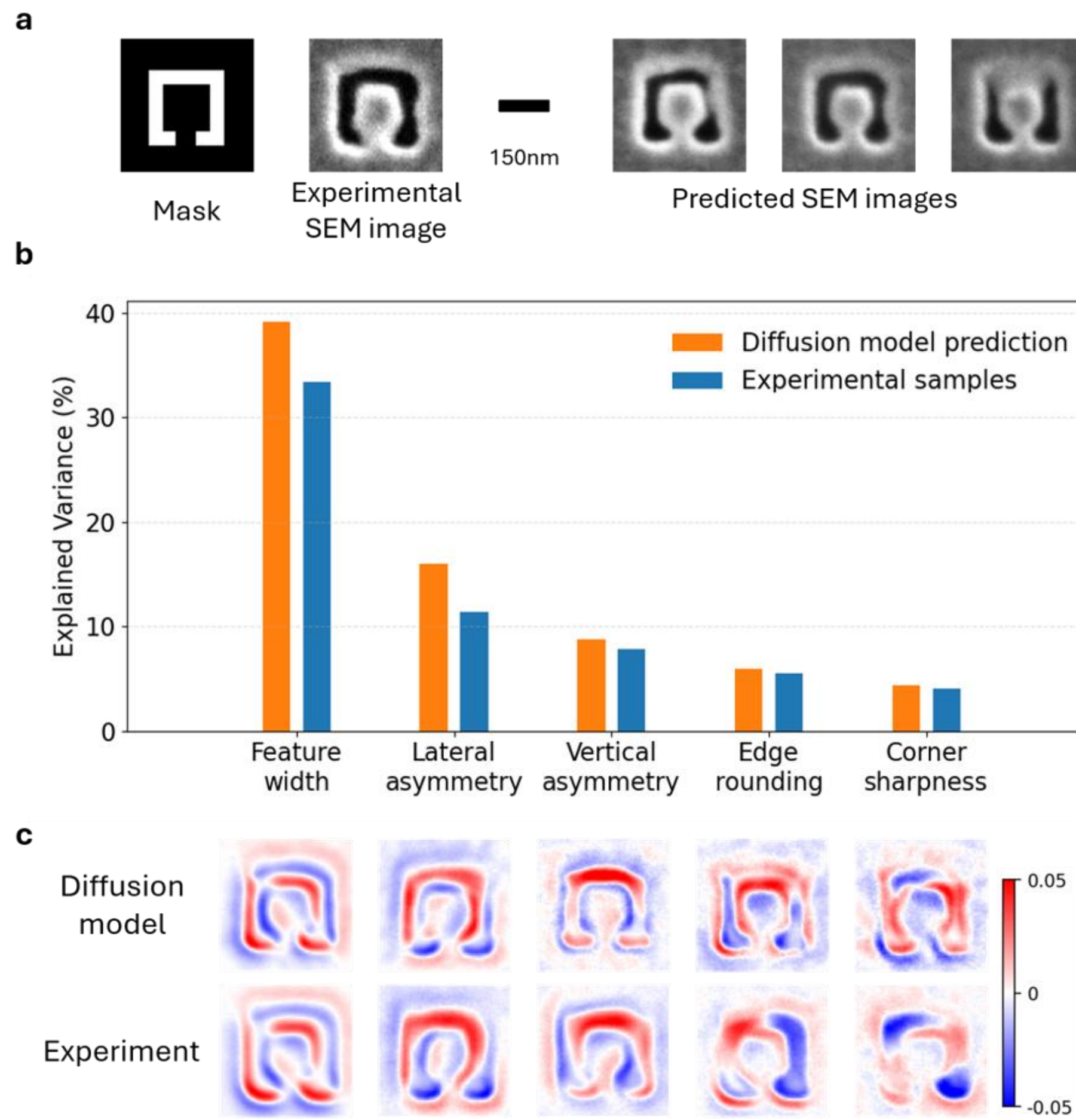

FIG 3. Principal component analysis of structural deformation modes diffusion model simulated vs. experimentally fabricated split-rings. (a) The split ring binary mask and corresponding representative experimental sample SEM images, and diffusion model-generated pseudo-SEM images. (b) Explained variance (EV) of the first five principal components of structural deformation in a decomposition of simulated pseudo-SEM images [orange] and fabricated sample SEM images [blue]. (c) Corresponding principal component eigenimages illustrating the spatial distributions of deformation modes. [The color scale represents dimensionless principal-component loadings, with positive/negative values denoting opposing directions of deviation from the mean structure – sign is arbitrarily assigned.]

Importantly, because the principal component decompositions were performed independently for the experimental and diffusion model-generated image sets, the close correspondence (which extends to the tenth component at EV ~$^1/_{100}$) arises entirely from similarity in the underlying structural variability, rather than from a shared (predefined) decomposition basis. It demonstrates that the diffusion model successfully captures the statistical hierarchy of geometric deformations governing stochastic fabrication variability.

## IV. CONCLUSIONS

In summary, we have developed and experimentally validated a conditional diffusion model for probabilistic prediction of focused ion beam milling outcomes directly from binary mask nanostructured material/device designs. Where other approaches to modelling nanofabrication processes are computationally and/or analytically demanding (e.g. requiring detailed knowledge of underlying physics/chemistry), and typically predict a single (average or most-likely) fabrication result, the diffusion model quickly generates a realistic ensemble of possible results, capturing and visualizing the statistical distribution of variations in fabrication outcome.

The model is efficiently and very effectively trained on a small set of random-pattern binary masks and corresponding fabricated sample SEM images. It subsequently generates pseudo-SEM images from previously unseen design masks, which accurately predict the statistical distributions of real sample geometries and feature dimensions. It correctly accounts for subtle milling performance anisotropy associated with ion beam raster scan orientation, the likelihood of defects associated with redeposition, and the hierarchy of structural deformation modes (identified by principal component analysis).

By explicitly simulating fabrication variability, the diffusion model provides a foundation for fabrication-aware design optimisation, realistic modelling of geometry-dependent dispersion in the optical/mechanical properties of nano/microstructured materials, and (thereby) tolerance-aware device and manufacturing process design. More broadly, this work demonstrates the potential of probabilistic generative models to connect nanostructural design with the reality of nanoscale fabrication, by predicting both the expected manufactured structure and the distribution of likely fabrication outcomes.

## ACKNOWLEDGMENTS

This work was supported by the Engineering and Physical Sciences Research Council (EPSRC), UK [grant numbers EP/U536866/1, EP/W028786/1].

## DATA AVAILABILITY

The data from this paper can be obtained from the University of Southampton ePrints research repository https://doi.org/10.5258/SOTON/D3997.